\documentclass[conference]{IEEEtran}
\IEEEoverridecommandlockouts
\usepackage{cite}
\usepackage{amsmath,amssymb,amsfonts}
\usepackage{algorithmic}
\usepackage{graphicx}
\usepackage{textcomp}
\usepackage{xcolor}
\usepackage{comment}
\usepackage{fancyvrb}
\usepackage{stfloats}

\usepackage{pgfplots} 
\usepgfplotslibrary{groupplots}
\pgfplotsset{compat=1.9}
\usepackage{tikz}
\usetikzlibrary{patterns} 

\def\BibTeX{{\rm B\kern-.05em{\sc i\kern-.025em b}\kern-.08em
    T\kern-.1667em\lower.7ex\hbox{E}\kern-.125emX}}
\begin{document}

\title{iPregel: Strategies to Deal with an Extreme Form of Irregularity in Vertex-Centric Graph Processing
\thanks{We thank the reviewers for their helpful feedback and suggestions. This research was supported by the UK Engineering and Physical Sciences Research Council (grant number EP/L01503X/1, CDT in Pervasive Parallelism)}
}

\author{\IEEEauthorblockN{Ludovic Anthony Richard Capelli}
\IEEEauthorblockA{\textit{School of Informatics} \\
\textit{The University of Edinburgh}\\
Edinburgh, United Kingdom \\
l.capelli@ed.ac.uk}
\and
\IEEEauthorblockN{Nick Brown}
\IEEEauthorblockA{\textit{Edinburgh Parallel Computing Centre} \\
\textit{The University of Edinburgh}\\
Edinburgh, United Kingdom \\
n.brown@epcc.ed.ac.uk}
\and
\IEEEauthorblockN{Jonathan Mark Bull}
\IEEEauthorblockA{\textit{Edinburgh Parallel Computing Centre} \\
\textit{The University of Edinburgh}\\
Edinburgh, United Kingdom \\
m.bull@epcc.ed.ac.uk}
}

\maketitle

\begin{abstract}
Over the last decade, the vertex-centric programming model has attracted significant attention in the world of graph processing, resulting in the emergence of a number of  vertex-centric frameworks. Its simple programming interface, where computation is expressed from a vertex point of view, offers both ease of programming to the user and inherent parallelism for the underlying framework to leverage.
However, vertex-centric programs represent an extreme form of irregularity, both inter and intra core. This is because they exhibit a variety of challenges from a workload that may greatly vary across supersteps, through fine-grain synchronisations, to memory accesses that are unpredictable both in terms of quantity and location.
\par
In this paper, we explore three optimisations which address these irregular challenges; a hybrid combiner carefully coupling lock-free and lock-based combinations, the partial externalisation of vertex structures to improve locality and the shift to an edge-centric representation of the workload.
The optimisations were integrated into the iPregel vertex-centric framework, enabling the evaluation of each optimisation in the context of graph processing across three general purpose benchmarks common in the vertex-centric community, each run on four publicly available graphs covering all orders of magnitude from a million to a billion edges. 
\par
The result of this work is a set of techniques which we believe not only provide a significant performance improvement in vertex-centric graph processing, but are also applicable more generally to irregular applications.
\end{abstract}


\section{Introduction}\label{Introduction}
Graphs have become an ubiquitous data structure due to their application throughout very many areas of technology. As such the development of graph processing algorithms is an important and active area of research, with growing interest in the development of programming models for this domain in order to support the expression of complicated algorithms. It follows that there is a growing interest in programming models for graph processing.
\par
In 2010, Google introduced the vertex-centric programming model through Pregel~\cite{Pregel}, where the user expresses the graph computation from a vertex perspective. In this model, vertices can communicate with each other by sending messages along their outgoing edges and have a mailbox for incoming messages. The execution first begins with all vertices active, a user-defined function is then applied to each one of these, and vertices can halt and become inactive as appropriate. Working in iterations, also called supersteps, once all vertices are processed for a specific step, message exchanges between vertices are performed and vertices that receive a message become, or remain, active. Once all message exchanges are performed, a new superstep begins: reapplying the user-function to all active vertices, then performing message exchanges before a new superstep begins. This iterative execution repeats until all vertices are inactive.
\par
Whilst this model provides significant programmability advantages, and exposes a large degree of latent parallelism, it suffers from numerous irregularities that impact performance. In fact, when it comes to common properties associated with irregular applications, the vertex-centric model exhibits many sources of irregularity:
\begin{itemize}
    \item \textbf{Fine-grain synchronisations}: the communications in vertex-centric programs take place at a vertex's mailbox level, hence any data-race protection must be implemented on a per-vertex basis.
    \item \textbf{Unpredictable memory access patterns}: broadcasting a message from a vertex to its neighbours means emitting a vertex-specific number of messages, aggravated by the power-law distribution of the numbers of neighbours per vertex. Also, the recipient vertices are unlikely to reside next to each other in memory, making these memory accesses unpredictable both in terms of quantity and location.
    \item \textbf{Load-imbalance}: the number of active vertices may drastically vary from a superstep to the next, and the numbers of neighbours may drastically vary too from a vertex to the next.
\end{itemize}
\par
Since the challenges faced in vertex-centric programs echo those of a larger class of irregular problems, optimisations that are successful for vertex-centric are likely to be applicable or adaptable to other heavily irregular applications such as social network analysis or graph databases for instance.
\par
Vertex-centric programs are inherently difficult to optimise because they tend to take the form of a short user provided source code, resulting in the fact that the underlying framework is provided with little information to leverage for performance. In the meantime, since programmability is the essence of vertex-centric, the performance optimisations designed must not degrade the programmability properties of the framework.
\par
Yet, many attempts to optimise the vertex-centric model have negatively impacted the ability to easily program this paradigm, highlighting the tension between performance optimisation and programmability~\cite{iPregelPARCO}.
By contrast, our approach integrates the preservation of vertex-centric programmability in its core. To that end, all the optimisations discussed in this paper are encapsulated within the iPregel framework, requiring no user source modification to take advantage of them.
\par
The main contributions presented in this paper can be summarised as follows:
\begin{itemize}
    \item \textbf{A hybrid combiner} designed to couple lock-free and lock-based interactions in order to efficiently handle fine-grain synchronisations.
    \item \textbf{The externalisation of vertex attributes} to better cope with unpredictable memory accesses by improving  the cache efficiency through the grouping of vertex attributes based on their access frequency.
    \item \textbf{An edge-centric workload representation} that improves load-imbalance while preserving both the vertex-centric paradigm and Pregel user interface.
\end{itemize}
\par
The rest of this paper is organised as follows: Section~\ref{Related work} depicts the context in which this research takes place. Sections~\ref{Fine-grain synchronisations}, \ref{Unpredictable memory access patterns} and \ref{Irregular workload} introduce the optimisations considered in this paper. Section~\ref{Experimental environment} describes the environment in which experiments were conducted while Section~\ref{Results} presents and analyses the results obtained. This paper then concludes in Section~\ref{Conclusions}; summarising the findings of this work as well as discussing potential future work directions.

\section{Related work}\label{Related work}
Over the last decade, the vertex-centric programming model has received great attention due to the programmability it offers to the user and the amount of parallelism inherently provided. There has been some attempts by the vertex-centric community to address the implications of irregular workloads present in graph processing algorithms.
\par
Bypassing the costly selection of active vertices has been addressed, whether it has to be done manually by the user such as GraphChi~\cite{GraphChi} or automatically thanks to the analysis of algorithmic patterns as introduced in early iPregel work~\cite{iPregel}. However, the efficient dispatching of active vertices to threads remains a challenge. Indeed, accurately evaluating the workload contained in these active vertices is key to an efficient workload dispatch. The common approach in vertex-centric frameworks consists in distributing an equal number of active vertices to each worker. However, this approach is sub-optimal due to the power-law degree distribution that typically underpins the graphs processed. This observation led to the development of PowerGraph~\cite{PowerGraph}, where the authors adopted a more edge-centric approach, but which resulted in an entirely new interface based on the scatter-apply-gather design instead of the original Pregel single user-defined function. As such, many of the abstraction and programmability benefits of the vertex-centric model were lost.
\par
The edge-centric approach was taken one step further by X-Stream\cite{XStream}, whose implementation and interface are entirely designed from an edge-centric perspective; exposing an edge-centric Gather-Apply-Scatter interface to the user. The underlying motivation for this design was to address the randomness of memory accesses, by reading the graph's edges sequentially. Nonetheless, such edge-centric frameworks, which by definition are no longer vertex-centric hence cannot provide the same benefits, demonstrate that addressing the vertex-centric challenges without sacrificing certain aspects of the actual programming model is not a trivial task.
\par
This is also illustrated in the implementations of optimisations addressing the fine-grain synchronisations required in vertex-centric programs during message exchange. Typically, either communications are redesigned as pull-based~\cite{GraphChi} so they are lock-free, or a semaphore is needed for each vertex mailbox where appropriate. An alternative approach for the latter is to implement the message combination as a compare-and-swap, as illustrated in Ligra~\cite{Ligra}. However, despite providing performance benefits, this design again reduces the level of abstraction and requires the end programmer to interact with the framework at a lower level, potentially requiring a rewriting of certain parts of their code, in addition to raising additional restrictions that will be discussed in details later in this paper.
\par
As described in this section, the application of optimisations for dealing with irregular workloads in vertex-centric frameworks typically results in the sacrifice of vertex-centric aspects, features or programmability. By contrast, in previous work~\cite{iPregelPARCO}, we demonstrated that vertex-centric optimisations could be designed without generating such unwanted side effects. In this paper, we continue this direction and focus on developing optimisations to cope with vertex-centric irregularities without sacrificing the actual vertex-centric programmability or requiring a user source code rewrite.

\section{Fine-grain synchronisations}\label{Fine-grain synchronisations}
Vertex-centric programs require each vertex mailbox to be protected against potential data-races. The write-interactions with that mailbox are achieved during combination, which makes combiners a key area for optimisations since any improvement in their design will have a direct impact on the performance observed. To implement them, two designs are available:
\begin{itemize}
    \item \textbf{lock}: a classic design where the sender vertex acquires the lock held on the recipient vertex, checks if that recipient vertex already received a message, and if so combines the existing message with the new one. Otherwise, it simply pushes the new message, before releasing the lock.
    \item \textbf{compare-and-swap}: a lock-free design where the sender vertex retrieves the existing message from the recipient vertex's mailbox, combines it with their own message and pushes the result back using a compare-and-swap operation that atomically checks if the value of the mailbox message is still identical to that read earlier. If so, it updates it with the new value and returns \emph{true}; otherwise, it means the value changed, which implies that another vertex updated this recipient's mailbox first, in which case the entire operation is repeated until it eventually succeeds.
\end{itemize}
\par
The second approach has the advantage of avoiding locks, thus resulting in a performance gain. However, it systematically combines the new message with the existing one, therefore relying on the assumption that mailboxes begin with a default message value that is neutral to the combination operation applied. For instance, in a combination operation that sums messages, vertices mailboxes would begin each superstep with a message value of 0. This need of a neutral value implies that either the user must be constrained to a set of predefined combination operations whose neutral values are hardcoded, or else the user must somehow declare the neutral value for the combination operation they write. In the Ligra version of PageRank, for instance, the combination operation is a sum (thus having the neutral value 0). For the user, this results in having to manually reset each vertex mailbox to 0 at the end of every superstep.
\par
The second drawback of a compare-and-swap design comes from the lack of a notion of empty mailboxes. Indeed, mailboxes always have a message, either representing the result of a combination, or being the neutral value by default. Therefore, a vertex knows it has received a message if the mailbox message value is different from the neutral value. Yet, in a scenario where the combination operation would result in the neutral value itself, the vertex would assume it has not received a message, while in fact it has. In vertex-centric programs, this can lead to incorrect programs since receiving messages is what reactivates inactive vertices.
\par
In order to make the best of both designs, that is; exploiting compare-and-swap while keeping the notion of an empty mailbox, as well as letting the user define any arbitrary combination operation, we designed a hybrid combiner that leverages lock-based and lock-free interactions with the recipient mailbox. Its implementation is provided in Figure~\ref{fig:hybridCombinerImplementation}, where vertex attributes have been shortened for brevity. In this example, \emph{ip\_combine} is the user-defined combination function, \emph{has\_msg\_next} is the flag indicating whether the vertex already received a message during this superstep and \emph{msg\_next} is the message itself (whose value is meaningful only if the flag is \emph{true}).
\begin{figure}
	\begin{Verbatim}[fontsize=\fontsize{8}{10}\selectfont\ttfamily,
	                 tabsize=2,
	                 frame=single,
	                 numbers=right]
void apply_cas(IP_VERTEX_TYPE* dst,
               IP_MESSAGE_TYPE msg){
  IP_MESSAGE_TYPE old_msg = dst->msg_next;
  IP_MESSAGE_TYPE new_msg = old_msg;
  ip_combine(&new_msg, msg);
  while(new_msg != old_msg &&
        !atomic_compare_exchange_strong(
          &dst->msg_next, &old_msg, new_msg)) {
    old_msg = dst->msg_next;
    new_msg = old_msg;
    ip_combine(&new_msg, msg);
} }

void ip_send_message(IP_VERTEX_ID_TYPE dst_id,
                     IP_MESSAGE_TYPE msg) {
  IP_VERTEX_TYPE* dst=ip_get_vertex_by_id(dst_id);
  if(dst->has_msg_next) apply_cas(dst, msg);
  else {
    ip_lock_acquire(&dst->lock);
    if(dst->has_msg_next) {
      ip_lock_release(&dst->lock);
      apply_cas(dst, msg);
    } else {
      dst->message_next = msg;
      dst->has_message_next = true;
      ip_lock_release(&dst->lock);
} } }
    \end{Verbatim}
	\caption{Implementation in iPregel of the hybrid combiner}
	\label{fig:hybridCombinerImplementation}
\end{figure}
\par
As shown in Figure~\ref{fig:hybridCombinerImplementation}, the hybrid combiner carefully couples lock-free and lock-based interactions. Correctness comes from the guarantee that if the \emph{has\_msg\_next} flag of a recipient vertex is \emph{true}, the value held in that vertex mailbox is set. Indeed, as soon as the \emph{has\_msg\_next} flag is \emph{true}, potential compare-and-swap combinations may happen concurrently on that vertex from other threads. Therefore, the value they will fetch from that recipient vertex mailbox must have been set by that time.
\par
To satisfy this guarantee, when a thread pushes the first message to a recipient mailbox, it stores the message (line 24) before setting the flag to \emph{true} (line 25). In addition, in order to avoid a potential out-of-order execution, a full memory barrier is required in-between. It is achieved by declaring the \emph{has\_msg\_next} flag as atomic, using C11 atomics, which implicitly enforces a sequentially consistent memory model. Without it, an out-of-order execution could result in a recipient vertex entering a state where its \emph{has\_msg\_next} flag is set to \emph{true} while not having its \emph{msg\_next} message set yet. Another thread meaning to push a message to that vertex mailbox would therefore check the flag, see it is \emph{true} and apply a compare-and-swap with the very message that is still unset. Finally, having the \emph{has\_msg\_next} flag as atomic implies that the read at line 17 and write at line 25 are atomic too; guaranteeing that a read cannot happen on a flag partially written to memory.
\par
In the rest of the hybrid combiner, threads check if the recipient vertex already has a message and if so they use a compare-and-swap combination, otherwise they acquire the lock. When the lock is acquired by a thread, it checks once again the recipient vertex flag in case, while it was waiting to acquire the lock, another thread that was holding that lock pushed the first message into that recipient mailbox. In this event, the recipient vertex now has a mailbox containing a set message therefore the thread can release the lock and use the compare-and-swap combination. Otherwise, it holds the lock and performs the first message push to that recipient vertex mailbox, making sure the store operations are issued in the order explained earlier.

\section{Unpredictable memory access patterns}\label{Unpredictable memory access patterns}
In vertex-centric programs, the irregularity in memory accesses is two-fold. First, the power-law distribution that typically underpins the graphs processed results in vertices having widely different numbers of neighbours. Second, the inherent irregular structure of graphs allows each vertex to be connected with any other arbitrary vertex. In other words, the data for neighbouring vertices may reside at any location in memory, thus unlikely to be contiguous with each other. As a consequence, when a vertex broadcasts a message to its neighbours, there are an arbitrary number of memory accesses to perform, each at an arbitrary memory location.
\par
These memory access patterns, although unpredictable, expose one regularity: the attributes accessed from these neighbours vertex structure. In iPregel, when using the pull-based version, the vertex structure contains among other attributes a \textit{flag} indicating if it has data to broadcast and a variable for the actual \textit{message}. The combination consists in iterating through neighbours, checking their \textit{flag} and fetching their \textit{message}.
\par
However, while never accessed during this combination process, the other vertex attributes are still loaded into cache along with the meaningful attributes sharing the same vertex structure. Nonetheless, this cache pollution can be minimised by reorganising the vertex structure; externalising the frequently accessed attributes into their own structure. In other words, one array would contain structures made of the \textit{flag} and \textit{message} attributes, while the other array would contain structures made of the rest of the vertex attributes. Therefore, such a design allows cache lines to be loaded only with useful attributes.

\section{Irregular workload}\label{Irregular workload}
A common irregularity that parallel programs face is that of load imbalance, where processes or threads have associated with them different amounts of work. Vertex-centric programs, where vertices can become inactive during execution and contain different numbers of edges, are therefore prone to load imbalance.
\subsection{Workload evaluation proxy}\label{Workload evaluation proxy}
Finding the right proxy to evaluate the workload is crucial since it lays down the foundations on which build more advanced strategies like load-balancing. Logically, implementations of the vertex-centric programming model represent workload in terms of vertices. Although this is accurate with regular data structures, the graphs processed by vertex-centric programs typically follow a power-law distribution, resulting in widely different number of neighbours per vertex. In addition, the runtime of typical vertex-centric programs is dominated by communications and not computation. While the latter is related to the number of vertices, the former depends on the number of edges. Based on this observation, our hypothesis was that the workload of a thread, which results in the number of combination operations performed and memory writes, or reads, is better expressed as being correlated to the number of outgoing, or incoming, neighbours.
\subsection{Work distribution}\label{Work distribution}
When parallelising a \emph{for} loop in OpenMP by using the \emph{for} construct, one can apply the \emph{schedule} clause, which is provided with a scheduling kind describing how chunks of iterations will be distributed to threads as well as an optional parameter specifying how many iterations make a chunk.
\par
One of the scheduling kinds provided by OpenMP is \emph{dynamic}, specifying that chunks of iterations will be distributed on a first-come-first-served basis. This allows threads that have been assigned lighter chunks to be assigned more chunks, thus improving load-balancing.
\par
To be compatible with this technique, the code must be within a \emph{for} loop whose iteration set distribution can be freely managed by OpenMP. This is compatible with all versions of iPregel that do not rely on the workload shift from vertex-centric to edge-centric described in Subsection~\ref{Workload evaluation proxy}. Indeed, the edge-centric workload negates the use of OpenMP dynamic scheduling because the workload is represented as edges and not vertices. The assigned chunks therefore represent workloads on a per-vertex basis instead of edge-centric one.

\section{Experimental environment}\label{Experimental environment}
This section describes the conditions and configurations in which the experiments presented in this paper were conducted.
\subsection{Computing environment}
Experiments are run on a standard compute node of an HPE 8600 cluster, set up with CentOS 7 Linux and containing two 2.1 GHz, 18-core Intel Xeon E5-2695 (Broadwell) series processors and 256GB of memory split in two 128GB non-uniform memory access regions, one local to each processor.
\par
The compilation is achieved by using the GCC compiler version 8.2.0 with OpenMP version 4.5. Compilation flags passed enable the support for C11 standard (\emph{-std=c11}) and level 3 optimisations (\emph{-O3}).
\subsection{Graph configurations}
Table~\ref{tab:graphs} lists the graphs processed in the experiments presented in this paper. All four are real-world graphs publicly available in the Stanford Network Analysis Project~\cite{SNAP} online collection. The smallest graph, the Database and Logic Programming Bibliography graph (DBLP), represents the eponymous computer science bibliography while LiveJournal, Orkut and Friendster are network graphs about blogging, social and gaming respectively. These graphs cover all orders of magnitude from a million to a billion edges and are undirected, meaning that the total number of directed edges is twice the amount presented.
\begin{table}
    \caption{Number of vertices and edges in the graphs selected for experiments}
	\begin{center}
		\setlength\tabcolsep{0.2cm}
		\begin{tabular}{lrrr}
			\hline
			\noalign{\smallskip}
			\multicolumn{1}{c}{Name}&\multicolumn{1}{c}{Vertex count}&\multicolumn{1}{c}{Edge count}\\
			\noalign{\smallskip}
			\hline
			\noalign{\smallskip}
			DBLP & 317,080 & 1,049,866 \\
			Live Journal & 4,036,538 & 34,681,189 \\
			Orkut & 3,072,441 & 117,185,083 \\
			Friendster & 65,608,366 & 1,806,067,135 \\
			\noalign{\smallskip}
			\hline
		\end{tabular}
	\end{center}
	\label{tab:graphs}
\end{table}
\subsection{Benchmarks}
Experiments presented in these paper are conducted on three benchmarks commonly used by the vertex-centric community.
\par
PageRank, or PR, is an iterative algorithm originally presented in~\cite{PageRank} that ranks webpages by evaluating their importance; taking into account a ratio between incoming and outgoing hyperlinks. In iPregel, PR is best implemented using the single-broadcast version, where communications are achieved by pulling messages from their sender's outbox.
\par
Connected Components, also abbreviated CC, locates in a graph all the subgraphs in which any two vertices are connected to each other by paths but connected to no vertex outside that subgraph. In iPregel, the CC benchmark is best implemented using the single-broadcast with selection bypass version, where in addition to leveraging pull-based communications, the framework keeps track of active vertices to better dispatch them to threads at every supersteps.
\par
Single-Source Shortest Path, or SSSP, consists in selecting a vertex and finding the shortest path from that vertex to each other vertex it can reach. Experiments presented in this paper use the unweighted version of SSSP, where all edges represent a distance of~1. In iPregel, SSSP is best implemented using the selection bypass version described above.
\par
Further details about each internal version of iPregel as well as a detailed analysis of the benchmarks can be found in~\cite{iPregel} and~\cite{iPregelPARCO}. As discussed in Section~\ref{Introduction}, we have designed the optimisations of Sections~\ref{Fine-grain synchronisations}, \ref{Unpredictable memory access patterns} and \ref{Irregular workload} in a manner that requires no modifications in user code. As such the versions of the benchmarks in~\cite{iPregel} and~\cite{iPregelPARCO} have remained unchanged in the experiments of this paper.

\section{Results}\label{Results}
\begin{table}
    \caption{Speed-ups obtained from each optimisation applied individually, then together, for each benchmark, using 32 threads, on all graphs ordered by ascending number of edges}
    \centering
    \begin{tabular}{rccccccc}
        \hline
        \noalign{\smallskip}
         & DBLP & Live Journal & Orkut & Friendster\\
        \noalign{\smallskip}
        \hline
        \noalign{\smallskip}
        \textbf{PR (10 iterations)} & & & & \\
        Baseline & 1.00 & 1.00 & 1.00 & 1.00\\
        Externalised structure & 1.31 & 1.27 & 1.51 & 1.13\\
        Edge-centric workload & 1.01 & 2.31 & 1.67 & 1.36\\
        Dynamic scheduling & 1.23 & 2.31 & 1.99 & 1.44\\
        Final & 1.61 & 3.14 & 3.07 & 1.63\\
        \noalign{\smallskip}
        \hline
        \noalign{\smallskip}
        \textbf{CC} & & & & \\
        Baseline & 1.00 & 1.00 & 1.00 & 1.00 \\
        Externalised structure & 1.58 & 1.66 & 1.47 & 1.65\\
        Edge-centric workload & 0.56 & 1.12 & 1.27 & 1.41\\
        Dynamic scheduling & 1.23 & 1.67 & 1.69 & 1.20\\
        Final & 2.05 & 2.96 & 2.41 & 2.12\\
        \noalign{\smallskip}
        \hline
        \noalign{\smallskip}
        \textbf{SSSP} & & & & \\
        Baseline & 1.00 & 1.00 & 1.00 & 1.00 \\
        Hybrid combiner & 1.01 & 1.12 & 2.35 & 4.07\\
        Externalised structure & 1.08 & 1.01 & 1.07 & 1.10\\
        Edge-centric workload & 0.91 & 0.87 & 1.28 & 1.29\\
        Dynamic scheduling & 1.11 & 1.33 & 1.55 & 1.69\\
        Final & 1.09 & 1.75 & 3.18 & 5.63\\
        \noalign{\smallskip}
        \hline
    \end{tabular}
    \label{tab:speedups}
\end{table}
The results presented in Table~\ref{tab:speedups} consist in applying every optimisation individually and calculating the speed-up obtained against the baseline version, before repeating the process with a version aggregating all applicable optimisations.
\subsection{Individual optimisations}
The results presented in Table~\ref{tab:speedups} show that the hybrid combiner improves the performance of SSSP on all graphs. It also proves to be the optimisation yielding both the biggest speed-up overall, up to 4.07 on Friendster, and on average, with a geometrical mean of 1.81. In addition, as the size of the graph increases, so does the speed-up. The reason for this is that the number of combinations depends on the number of edges, the benefit of improving the combination therefore grows along with the number of combinations generated. These results demonstrate that efficiently handling fine-grain synchronisations does not have to sacrifice programmability.
\par
Similarly, the externalised structure optimisation is beneficial in all graph-benchmark pairs tested, with a speed-up of 1.30 on average. The results in Table~\ref{tab:speedups} show that externalising vertex attributes generates the best speed-ups for CC and the worst ones for SSSP. The explanation is twofold, firstly, PR and CC benefit more because they rely on iPregel versions that use pull-based communications that are lock-free by design. As a consequence, the memory accesses performed during the communications are not interleaved with lock acquisition or release, which reduces further the number of vertex attributes that are frequently accessed. Secondly, PR and CC rely on different algorithms; the one underpinning PR has one loop that can leverage structure externalisation while the one for CC has two. The overall benefit obtained on CC is therefore greater since it can leverage this optimisation in two parts of the code. Overall, structure externalisation therefore demonstrates that heavily irregular memory access patterns may exhibit certain regular aspects when analysed from a different angle, which can be leveraged for performance gains.
\par
The timings reported in Table~\ref{tab:speedups} also indicate that shifting to the edge-centric workload proves to be beneficial in 75\% of the experiments; yielding a speed-up of 1.19 on average. The extremes are observed for PR on Live Journal with a speed-up of 2.31 and Connected Components on DBLP with only 0.56. In fact, the edge-centric workload representation performs better on PR than on any of the two other benchmarks. The reason for this is that CC and SSSP rely on an iPregel implementation leveraging the \emph{selection bypass} optimisation introduced in~\cite{iPregel}, which helps cope with variable number of active vertices but requires the edge-centric workload distribution to be recalculated at every superstep, therefore increasing the total overhead.
\par
The OpenMP \emph{dynamic} scheduling is the fourth optimisation explored in these experiments. The results presented in Table~\ref{tab:speedups} are obtained with an empirically determined chunk size of 256. Unlike the edge-centric optimisation, the \emph{dynamic} scheduling turns out to improve the performance in all experiments; resulting in speed-ups between 1.11 and 2.31. With an average speed-up of 1.50, the first-come-first-served dispatch pattern proves to be an efficient load-balancing strategy.
\subsection{Aggregated optimisations}
The versions referred to as "final" combine all optimisations applicable. They do not include the hybrid combiner in the case of PR and CC since these benchmarks rely on iPregel versions that are lock-free by design. Also, because of the incompatibility between the edge-centric shift and dynamic scheduling, the former was excluded in favour of the better performing dynamic scheduling.
\par
For PR and CC, the smallest and biggest graphs benefit the least. Live Journal and Orkut on the other hand show better speed-ups, with Live Journal benefiting the most. In the case of PR, the speed-ups obtained range from 1.61 up to 3.14. For CC, the speed-up range is less spread; reaching a maximum of 2.96 but never going below 2.05.
\par
For SSSP, the speed-up pattern exhibited is different: the bigger the graph, the higher the speed-up as Table~\ref{tab:speedups} shows. This can be explained by the presence of the hybrid combiner, which provides such a speed-up pattern, in addition to the dynamic scheduling also exhibiting that pattern for SSSP in Table~\ref{tab:speedups}. Starting at 1.09 on the smallest graph, the speed-up obtained increases until reaching 5.63 on the biggest graph.
\par
Overall, when fixing the number of threads at 32, the optimised versions prove to be beneficial for all benchmarks on all graphs tested. On average, the optimised versions cut almost two thirds (59\%) of the runtime measured, with extremes cases observed of 8\% and 82\%.

\section{Conclusions and future work}\label{Conclusions}
In this paper we explored multiple optimisations to address the irregular challenges inherent to vertex-centric programs.
\par
The fine-grain synchronisations that underpin message combinations presented in Section~\ref{Fine-grain synchronisations} was the first optimisation investigated. By carefully coupling compare-and-swap and lock-based operations, we designed a hybrid combiner that, while being transparent to the user hence preserving vertex-centric programmability, can drop the runtime by up to 75\% as shown in Table~\ref{tab:speedups}.
\par
In Section~\ref{Unpredictable memory access patterns}, we analysed the unpredictable memory access patterns from a different perspective and found that, although one cannot know which structure will be accessed next, one knows which structure attribute will. We therefore exploited this characteristic to redesign vertex structures for cache efficiency by externalising vertex attributes and grouping them according to their access frequency. This optimisation permitted to decrease the runtime measured by up to 40\% as shown in Table~\ref{tab:speedups}.
\par
Our approach for the third challenge, the load imbalance presented in Section~\ref{Irregular workload}, was to better evaluate the workload by representing it with an edge-centric metric while preserving the user interface. Table~\ref{tab:speedups} reports that this shift, although beneficial in 75\% of the tests and providing a runtime reduction of up to 57\%, proved to be less productive than the OpenMP \emph{dynamic} scheduling which never resulted in performance degradation while equalling the maximum runtime gain of edge-centric.
\par
Overall, the experiments conducted in this work show that the optimisations considered yield performance benefits in 37 out of the 40 graph-benchmark pairs tested. When combined, these successful optimisations proved to yield performance benefits in all graph-benchmark pairs tested as shows Table~\ref{tab:speedups}. This demonstrates that although the vertex-centric model exhibits many sources of irregularity, they can be efficiently addressed, reducing the runtime measured by up to 82\%.
\par
Future directions for this work could include the integration of work-stealing in the edge-centric workload, for example by designing an affinity schedule tailored for edge-centric. Another direction could be that of incrementalisation~\cite{Incrementalisation}; an optimisation area under-explored in vertex-centric but which could unlock a new level of performance. Finally, focusing on distributed-memory architectures would raise new challenges, calling for new optimisations to be designed.

\bibliographystyle{IEEEtran}
\bibliography{vertex_centric.bib}

\end{document}